\documentclass[final,5p,times,twocolumn,sort&compress]{elsarticle}
\usepackage{amsmath,amssymb,amsfonts,graphicx,slashed,citesort,dsfont}
\usepackage[usenames]{color}
\usepackage{ulem}
\usepackage{nicefrac} 
\usepackage[figuresright]{rotating}
\usepackage[rightcaption]{sidecap}
\usepackage{auto-pst-pdf}
\usepackage{psfrag}
\usepackage{lipsum}

\newcommand{\be}{\begin{equation}}
\newcommand{\ee}{\end{equation}}
\newcommand{\ba}{\begin{eqnarray}}
\newcommand{\ea}{\end{eqnarray}}
\let\oldhat\hat
\renewcommand{\vec}[1]{\boldsymbol{#1}}
\renewcommand{\hat}[1]{\oldhat{\boldsymbol{#1}}}
\allowdisplaybreaks[1]
\begin{document}

\begin{frontmatter}
\title{Regularization Methods for Nuclear Lattice Effective Field Theory}

\address[Bonn]{Helmholtz-Institut f\"ur Strahlen- und Kernphysik, Universit\"at Bonn, D--53115 Bonn, Germany}
\address[Juelich]{Institut fur Kernphysik, Institute for Advanced Simulation, JARA-HPC and J\"ulich Center for Hadron Physics,\\ 
Forschungszentrum J\"ulich, D--52425 J\"ulich, Germany}
\address[NCSU]{Department of Physics, North Carolina State University, Raleigh, North Carolina 27695, USA}
\author[Bonn]{Nico~Klein}
\author[NCSU]{Dean~Lee}
\author[Bonn]{Weitao~Liu}
\author[Bonn,Juelich]{Ulf-G.~Mei{\ss}ner}

\begin{abstract} 
We investigate Nuclear Lattice Effective Field Theory for the 
two-body system for several lattice spacings at lowest order in the
pionless as well as in the pionful theory.
We discuss issues of regularizations and predictions for the 
effective range expansion. In the pionless case, a simple Gaussian
smearing allows to demonstrate lattice spacing independence over
a wide range of lattice spacings. We show that regularization methods 
known from the continuum formulation are necessary as well as 
feasible for the pionful approach.
\end{abstract}


\end{frontmatter}

\section{Introduction}

The Nuclear Lattice Effective Field Theory (EFT) method \cite{Lee:2008fa} 
has led to impressive progress in the last decade and it has been applied to 
few- and many-body-systems successfully, for reviews see e.g. 
Refs.~\cite{Epelbaum:2008ga,ugmNPN}. The lattice spacing serves as
a natural UV regulator for the theory, as for a given value of $a$
the maximal momentum is $p_{\rm max} = \pi/a$. Although these calculations 
give a quite good description for the phase shifts, energy levels, etc., almost 
all calculations have been done for a fixed lattice spacing $a \simeq 2\,$fm,
corresponding to a soft momentum cut-off of about 300~MeV. This allows one
to treat all corrections beyond leading order (LO) in perturbation theory.
However, the cut-off dependence or lattice spacing dependence has not been analyzed 
systematically and there are still some problems in the two-nucleon system
like the relatively poor description of the $^3$S$_1$-$^3$D$_1$ mixing angle \cite{Borasoy:2007vy}. 
Further, such soft potentials seem to lead to some overbinding in medium-mass
nuclei, as discussed in Ref.~\cite{Lahde:2013uqa}.
Also, it has been shown that the leading order four-nucleon contact interactions need
to be smeared to avoid a cluster instability when four nucleons reside on one lattice site
 \cite{Borasoy:2006qn}. One might argue that the extension of such smearing methods also to the pion 
exchange contributions leads to a natural regularization of the lattice
EFT, allowing to vary the lattice spacing freely but using an explicit momentum
cut-off in the spirit of the work of Ref.~\cite{Montvay:2012zz}. More precisely, this 
inherent physical cut-off was implemented by formulating the lattice action in terms 
of blocked fields.

In this paper, we will focus on the 
neutron-proton two-body system at lowest order and discuss the lattice 
spacing dependence systematically. In addition, we discuss the necessity 
of regularizing the one-pion-exchange potential and provide a method that
goes beyond smearing and is borrowed from continuum calculations, which 
leads to the lattice spacing independence of observables for a broad range in $a$, see Ref.~\cite{Epelbaum:2014efa}.

While most of the calculations solve the transfer matrix using Monte Carlo 
methods or the Lanczos method for small eigenvalues of large sparse matrices, 
we use here the Hamiltonian formalism and solve it with the Lanczos method. 
Using this approach we can eliminate the discretization in the time direction 
and we have to consider only the variation in the position space discretization. 
In the following, all expressions are given in lattice units and one has to 
multiply the lattice results by the appropriate power of the lattice spacing $a$ 
to get the physical values. Note also that we show simulations for various large
enough volumes so that L{\"u}scher's finite volume formulas are sufficient for the infinite volume extraction and we can entirely focus on the remaining dependence on
the lattice spacing. 

In what follows, we will first display the necessary formalism to calculate the
neutron-proton system to lowest order on the lattice. It is important to already improve
the free Hamiltonian so as to be as close as possible to the free non-relativistic
dispersion relation. At very low energies,
one can consider the theory with contact interactions only, the so-called pionless
theory. As we will show, the smearing of the contact interactions can be used as
a regulator, leading to regulator-independent results for a broad range of values 
of the lattice spacing $a$. Matters are different in the pionful theory, which to LO
consists of two four-nucleon contact interaction and the long-ranged static 
one-pion-exchange potential (OPEP). As will be shown, combining the smearing of the
contact interactions with a position-space regularization of the OPEP will again
lead to results largely independent of $a$ for the physically sensible range of lattice spacing. 
Hence, one could use this modified leading-order approach to improve the current auxiliary field Monte Carlo simulations in 
Nuclear Lattice EFT.
In principle, now it is possible to consider the continuum limit $a\to 0$,
however, we refrain from doing that here, as it is sufficient to demonstrate lattice 
spacing independence for a physically relevant range of $a$.

\section{The lattice Hamiltonian}

To set the stage and to introduce our notations, we first discuss the free
Hamiltonian. Its discretized form reads
\begin{equation}
\begin{split}
&H_{\mathrm{free}}=\frac{1}{2m_N}\sum_{\vec{n},i,j}\sum_{\hat{l}}\left\{2\omega_0a_{i,j}^\dagger\left(\vec{n}\right) a_{i,j}^{}\left(\vec{n} \right)+\sum_{k=1}^3 (-1)^k\right.\\
&\left.\times \,\, \omega_k   \left[a_{i,j}^\dagger\left(\vec{n}\right)a_{i,j}^{} \left(\vec{n}+k\hat{l}\right)+a_{i,j}^\dagger \left(\vec{n}\right)a_{i,j}^{}\left(\vec{n}-k\hat{l}\right) \right]\right\}\, .\\
\end{split}
\end{equation}
Here, $a_{i,j}^{}, a^\dagger_{i,j}$ are the fermionic annihilation and creation operators 
with spin and isospin indices $i,j$, respectively, $m_N =(m_p+m_n)/2$ 
denotes the nucleon mass and $\hat{l}$ 
is a unit vector in spatial direction. The summation is over all lattice points 
$\vec{n}$ on the $L^3$ lattice. We use a stretched $\mathcal{O}(a^m)$-improved action 
and its coefficients $\omega_k$ are summarized in Tab.~\ref{tab:dercoeffa4N}, see
e.g. Ref.~\cite{Lee:2007jd,Lu:2014xfa}. $m$ indicates the number of  hopping points beyond next-neighbor interaction used in the Laplacian discretization in each spatial direction 
and we use $m=4$ throughout this paper. The stretching factor $N$ is introduced to minimize the errors arising from 
the discretized dispersion relations on the lattice especially for large momenta 
where the discretization does not approximate the continuum relation 
$E=\vec{p}^{\,2}/(2m_N)$ anymore. While there is some arbitrariness on the exact choice of $N$ depending on the values of the respective momentum, $N=3.5$ is a sensible choice.

\renewcommand{\arraystretch}{1.15}
\begin{table}[!htb]
\centering
\begin{tabular}{|c|c||c|c|}
\hline
&$\mathcal{O}(a^4)$&&$\mathcal{O}(a^2)$\\
\hline
$\omega_0$&$N \cdot \frac{1}{9} + \frac{49}{36}$&$o_0$&$0$\\
$\omega_1$&$N \cdot \frac{1}{6} + \frac{3}{2}$&$o_1$&$\frac{4}{3}$\\
$\omega_2$&$N \cdot \frac{1}{15} +\frac{3}{20}$&$o_2$& $\frac{1}{6}$\\
$\omega_3$&$N \cdot\frac{1}{90} + \frac{1}{90}$&$o_3$&$0$\\
\hline
\end{tabular}
\caption{Coefficients for the lattice discretization of the Laplacian, the dispersion relation and momentum components depending on the stretching factor $N$.}
\label{tab:dercoeffa4N}
\end{table}
\renewcommand{\arraystretch}{1.0}

The interaction potential consists of two/three terms in the pionless/pionful
theory at lowest order. 
The contact interaction consists of two terms which can be chosen as 
\begin{equation}
H_{\mathrm{cont}}= \frac{1}{2}
\sum_{\vec{n}}\left[ {C} \rho^{a^\dagger,a}\left(\vec{n}\right)\rho^{a^\dagger,a}\left(\vec{n}\right) 
+ {C_I}\sum_I \rho_I^{a^\dagger,a}\left(\vec{n}\right)\rho_I^{a^\dagger,a}\left(\vec{n}\right)\right]~,
\end{equation}
where the terms are summed over all lattice sites $\vec{n}$ and the isospin index $I=1,2,3$. These terms appear in both versions of the EFT considered here. In the pionful theory,
one has in addition the one-pion-exchange potential 
\begin{equation}
\begin{split}
H_{\mathrm{OPE}}=-\frac{g_A^2}{8F_\pi^2}\sum_{S_1,S_2,I} \sum_{\vec{n}_1,\vec{n}_2}&G_{S_1S_2}\left(\vec{n}_1-\vec{n}_2\right)\\ &\times\rho_{S_1,I}^{a^\dagger,a}\left(\vec{n}_1\right) \rho_{S_2,I}^{a^\dagger,a}\left(\vec{n}_2\right)~,
\end{split}
\label{eq:HamOPE}
\end{equation}
with $g_A$ the axial-vector coupling constant and $F_\pi$ the pion decay constant. $S_1$, $S_2$ are the respective spin indices which run from 1 to 3.
The corresponding lattice density operators read
\begin{align}
\rho^{a^\dagger,a}\left(\vec{n}\right)&=a^\dagger_{i,j} \left(\vec{n}\right) a_{i,j}\left(\vec{n}\right),\\
\rho^{a^\dagger,a}_I\left(\vec{n}\right)&=a^\dagger_{i,j} \left(\vec{n}\right) \tau_{I,jj^\prime}a_{i,j}\left(\vec{n}\right),\\
\rho^{a^\dagger,a}_{S,I}\left(\vec{n}\right)&=a^\dagger_{i,j} \left(\vec{n}\right) \sigma_{S,ii^\prime}\tau_{I,jj^\prime}a_{i,j}\left(\vec{n}\right),
\end{align}
and $G_{S_1S_2}(\vec{n})$ represents the pion propagator times the pion-nucleon vertex 
and is defined as
\begin{equation}
G_{S_1S_2}\left(\vec{n}\right)=\frac{1}{L^3}\displaystyle\sum_{\vec{p}}\frac{\exp\left(-i \vec{p}\cdot\vec{n}\right)\nu\left(p_{S_1}\right)\nu\left(p_{S_2}\right)}{1+ \frac{2}{q_\pi}\displaystyle\sum_{k=1}^3\sum_{l}(-1)^k\cos\left(kp_l\right)}
\label{eq:GS1S2}
\end{equation}
with $q_\pi=M_\pi^2+6\omega_0$. $\nu(p_{S_1})$, $\nu(p_{S_2})$ are the discretized momentum components of the first and second pion field which yields $\nu(p_l)= o_1\sin(p_l) \allowbreak - \allowbreak o_2\sin(2p_l) = p_l(1+\mathcal{O}(p_l^4))$ with the coffecients summarized in Tab.~\ref{tab:dercoeffa4N}. We only use a $\mathcal{O}(a^2)$ discretization, because we do not want to expand the respective interaction too much. A further improved momentum approximation is linked to a further expanded derivative in position space including more interactions at distinct lattice points and the locality of the pion-nucleon interaction is lost.
These momenta arise from the pion field derivative in the pion nucleon Lagrangian $\mathcal{L}_{\pi N}=-g_A/(2F_\pi)N^\dagger\vec{\tau}\cdot(\vec{\sigma}\cdot\vec{\nabla})\vec{\pi}N$.
To arrive at Eq.~(\ref{eq:GS1S2}), we note that 
the pion propagator is derived from the discrete action for instantaneous pions which takes the form~\cite{Borasoy:2006qn}
\begin{equation}
\begin{split}
S_{\pi\pi}\left(\pi_I\right)=&\left(\frac{m_\pi^2}{2}+3\omega_0\right) \sum_{\vec{n}}\pi_I\left(\vec{n}\right)\pi_I\left(\vec{n}\right)\\ &+\sum_{\vec{n},\hat{l}, k}\left(-1\right)^k\omega_k\pi_I\left(\vec{n}\right)\pi_I\left(\vec{n}+k\hat{l}\right)~.
\end{split}
\end{equation}
This is reparametrized by $\pi_I^\prime(\vec{n})=\sqrt{q_\pi}\pi_I(\vec{n})$. Finally, the new pion action reads
\be
\begin{split}
S_{\pi\pi}\left(\pi_I^\prime\right)=&\frac{1}{2}\sum_{\vec{n}}\pi_I^\prime\left(\vec{n}\right) \pi_I^\prime\left(\vec{n}\right)\\
&+\frac{1}{q_\pi}\sum_{\vec{n},\hat{l},k}\left(-1\right)^k\omega_k \pi^\prime\left(\vec{n}\right)\pi_I^\prime\left(\vec{n}+k\hat{l}\right)
\end{split}
\ee
and the respective pion propagator reads
\begin{equation}
D_\pi\left(\vec{p}\right)=\left[{1+\frac{2}{q_\pi}\displaystyle\sum_{k=1}^3\sum_{l}(-1)^k\cos\left(kp_l\right)}\right]^{-1}~.
\end{equation}
Furthermore, we introduce a Gaussian smearing 
\begin{equation}
f\left(\vec{p}\right)=\frac{1}{f_0}\exp\left[-b\frac{\tilde{\nu}\left(\vec{p}\right)}{2}\right]
\label{eq:smearing}
\end{equation}
with a stretched $\mathcal{O}(a^4)$ improved discretization 
\begin{equation}
\tilde{\nu}(\vec{p}) = 2\sum_{k=0}^3 \allowbreak \sum_{l=1}^3(-1)^k \omega_k\cos(kp_l)=\vec{p}^2\left[1+\mathcal{O}\left(p^6\right)\right]~,
\end{equation} 
where the error estimation is valid for $N=0$ and the 
coefficients given in Tab.~\ref{tab:dercoeffa4N}. $f_0$ is necessary for normalization reasons and is given by 
$f_0=(1/L^3)\sum_{\vec{p}}\exp[-b\tilde{\nu}(\vec{p})/2]$. This smearing modifies the contact interaction in momentum space according to
\be
\begin{split}
H_{\mathrm{cont}}= \frac{1}{2}
\sum_{\vec{p}}f\left(\vec{p^{\,}}\right)&\left[ {C} \rho^{a^\dagger,a}\left(\vec{p}\right)\rho^{a^\dagger,a}\left(\vec{p}
\right)\right.\\ 
&\left.+ {C_I}\sum_I \rho_I^{a^\dagger,a}\left(\vec{p}\right)\rho_I^{a^\dagger,a}\left(\vec{p}\right)\right]~.
\end{split}
\label{eq:contIntmod}
\ee
Such a smearing was introduced in Ref.~\cite{Borasoy:2006qn} to reduce the 
effect of high momentum contributions and remove the clustering 
instability of the contact interaction. As the leading order (LO) contribution is
iterated to all orders, such a smearing sums up some of the higher order
corrections. All other higher order corrections are then treated in perturbation
theory (as long as the lattice spacing is sufficiently large). Here, we concentrate
on the lowest order and leave the discussion of the treatment of the
higher order effects to a later publication. In any case, the smearing of the
LO contact terms is a useful tool to improve the description of the S-waves 
in a very efficient way without including all higher-order terms in a chiral 
counting scheme. In general, it is not necessary that the smeared contact interactions for the $^1S_0$- and $^3S_1$-channel have the same smearing function, $f(\vec{p})$. However, this is important for auxiliary-field Monte Carlo lattice simulations. Without the same smearing function for the two channels, the Monte Carlo simulation would have a far bigger problem with sign oscillations. For this reason we consider only one smearing function for both channels. The fact that we can approximately describe the effective ranges for both channels using only one smearing function can be viewed as a feature of the approximate SU(4) Wigner symmetry of the two-nucleon interaction.

In continuum chiral EFT it is necessary to regularize the one-pion-exchange potential due to its singularity at very small distances. We will show that such a singularity also appears for small lattice spacings and we will regularize it in position space as suggested in Ref.~\cite{Epelbaum:2014efa}
\begin{equation}
\tilde{f}\left(\vec{r}\right)
=\left[1-\exp\left(-\frac{\vec{r}^{\,2}}{2b}\right)\right]^n~,
\label{eq:possmearing}
\end{equation}
where the denominator is motivated by $\mathcal{F}_q\{\exp[-r^2/(2b)]\}\propto \exp(-bq^2/2)$ and $n$ is a free parameter.

In the following we study the two-nucleon system for different lattice spacings. 
Therefore we use the finite-volume formulas for the binding energy and 
the effective range expansion given by L\"uscher \cite{Luscher:1985dn,Luscher:1986pf}
\begin{align}
E_B(L)-E_B^\infty&=-24\lvert A\rvert^2 
\frac{\exp\left(-\sqrt{-2\mu E_B^\infty}L\right)}{2\mu L}, \label{eq:EBL}\\
p\cot\delta_0(p)&=
\frac{1}{\pi L} S(\eta),\quad\eta=\left(\frac{Lp}{2\pi}\right)^2~, \label{eq:pcotdL}
\end{align}
with $A$ the normalization constant of the wave function for large distances
and $\mu = m_pm_n/(m_p+m_n)$  the reduced mass.
The energy eigenvalues of the system are linked to the momentum squared by 
$E=p^2/(2\mu)$ and $S(\eta)$ is the zeta function. Its expansion for small 
$\eta$ is given by
\begin{equation}
S\left(\eta\right)=-\frac{1}{\eta}+S_0+S_1\eta^1+S_2\eta^2+S_3\eta^3+\ldots
\end{equation}
and the coefficients $S_i$ can be found in \cite{Borasoy:2006qn}.
We do calculations at volumes large enough that L{\"u}scher's finite volume formulas are sufficient for infinite volume extraction with negligible residual finite volume dependence so we can entirely focus on the remaining dependence on the lattice spacing.  For the scattering state calculations, the finite volume corrections are of size $\exp(-L/R)$, where $L$ is the box length and $R$ is the range of the interactions \cite{Luscher:1985dn}.  For the bound state calculations, the finite-volume formula in Eq.~(\ref{eq:EBL}) captures the leading $\exp(-\sqrt{-2\mu E^\infty_B}L)$, but there are also smaller corrections of size 
$\exp(-\sqrt{2}\sqrt{-2\mu E^\infty_B}L)$~\cite{Luscher:1986pf}.
For S-waves, the effective range expansion is $p \cot\delta_0(p) \approx -1/a_s+(r_e/2)p^2$ with $a_s$ the scattering length and $r_e$ the effective range. Hence, we match our finite volume results to the infinite volume effective 
range parameters and the deuteron binding energy, 
$a_{^{\,1}S_0} = -23.76(1)$~fm, $r_{^{\,1}S_0} = 2.75(5)$~fm,
 and $E_b =-2.224575(9)$~MeV.


In the fits,  different lattice volumes are used to produce enough data 
points to make the respective fit using 
Eqs.~(\ref{eq:EBL},\ref{eq:pcotdL}) and predict $r_{^{\,3}S_1}$ as well as 
$a_{^{\,3}S_1}$. Finally, one can compare the prediction to the  experimental values given by $a_{^{\,3}S_1} = 5.424(4)$~fm and $r_{^{\,3}S_1} = 1.759(5)$~fm. In the following we will repeat this procedure for lattice 
spacings between $a^{-1}=100$~MeV and $a^{-1}=400$~MeV, respectively, that is approximately between $2$~fm and $0.5$~fm in the pionless theory and 
for $a$ between $0.3$ and $2.0\,$fm the EFT with pions.

\section{The pionless theory}

\renewcommand{\arraystretch}{1.05}
\begin{SCtable*}[20][htb]
\begin{tabular}{|c|c|c|c|c|c|c|}
\hline
$a^{-1}$~[MeV]&$a$~[fm]&$C_{^3S_1}$~[MeV]&$C_{^1S_0}$~[MeV]&$b$~[MeV$^{-2}$]& $a_{^{\,3}S_1}$~[fm]&$r_{^{\,3}S_1}$~[fm]\\
\hline
100	&1.97& $-46.66$&$-28.18$&$3.83\cdot10^{-5}$&$5.611(1)$&$ 2.029(1)$\\  
130	&1.52& $-50.71$&$-30.16$&$3.93\cdot10^{-5}$&$5.630(1)$&$ 2.047(1)$\\
160	&1.23& $-50.26$&$-29.84$&$4.04\cdot10^{-5}$&$5.624(1)$&$ 2.034(1)$\\
200	&0.99& $-49.89$&$-29.70$&$4.13\cdot10^{-5}$&$5.614(1)$&$ 2.017(1)$\\
240	&0.82& $-50.17$&$-29.92$&$4.19\cdot10^{-5}$&$5.607(1)$&$ 2.008(1)$\\
300	&0.66& $-50.63$&$-30.24$&$4.23\cdot10^{-5}$&$5.602(1)$&$ 2.005(1)$\\
350	&0.56& $-50.85$&$-30.38$&$4.25\cdot10^{-5}$&$5.601(1)$&$ 2.008(1)$\\
400	&0.49& $-50.91$&$-30.43$&$4.27\cdot10^{-5}$&$5.605(1)$&$ 2.015(1)$\\
\hline
\end{tabular}
\label{tab:LECsmnopion}
\caption{Fit results in the Gaussian-smeared pionless EFT case for 
$L=34,36,38$ and $N=3.5$. The values for $a_{^{\,3}S_1}$ and $r_{^{\,3}S_1}$ are predictions (modulo the consistency condition Eq.~(\ref{eq:EBar-relation})).}
\end{SCtable*}
\renewcommand{\arraystretch}{1.0}

\begin{figure}[!htb]
\psfrag{C_3S1 [MeV^1]}[][]{\rotatebox{180}{\scriptsize$C_{^3S_1}$~[MeV]}}
\psfrag{C_1S0 [MeV^1]}[][]{\rotatebox{180}{\scriptsize$C_{^1S_0}$~[MeV]}}
\psfrag{a [fm]}[][]{\scriptsize$a[\mathrm{fm}]$}
\psfrag{b [MeV^-2]}[][]{\rotatebox{180}{\scriptsize$b$~[$10^{-5}$~MeV$^{-2}$]}}
\psfrag{L=14,17,20; N=3.5}[][]{\scriptsize L=14,17,20; N=3.5}
\psfrag{L=34,36,38; N=3.5}[][]{\scriptsize L=34,36,38; N=3.5}
\psfrag{L=14,17,20; N=0.0}[][]{\scriptsize L=14,17,20; N=0.0}
\psfrag{L=34,36,38; N=0.0}[][]{\scriptsize L=34,36,38; N=0.0}
\psfrag{L=14,17,20}[][]{\scriptsize L=14,17,20}
\psfrag{L=34,36,38}[][]{\scriptsize L=34,36,38}
\psfrag{0.0}{\scriptsize 0}
\psfrag{-28}{\scriptsize -28}
\psfrag{-29}{\scriptsize -29}
\psfrag{-30}{\scriptsize -30}
\psfrag{-31}{\scriptsize -31}
\psfrag{-32}{\scriptsize -32}
\psfrag{-33}{\scriptsize -33}
\psfrag{-46}{\scriptsize -46}
\psfrag{-48}{\scriptsize -48}
\psfrag{-50}{\scriptsize -50}
\psfrag{-52}{\scriptsize -52}
\psfrag{-54}{\scriptsize -54}
\psfrag{0}{\scriptsize 0}
\psfrag{0.5}{\scriptsize 0.5}
\psfrag{1.0}{\scriptsize 1.0}
\psfrag{1.5}{\scriptsize 1.5}
\psfrag{3.5}{\scriptsize 3.5}
\psfrag{4.0}{\scriptsize 4.0}
\psfrag{4.5}{\scriptsize 4.5}
\psfrag{5.0}{\scriptsize 5.0}
\psfrag{5.5}{\scriptsize 5.5}
\includegraphics[width=0.48\textwidth]{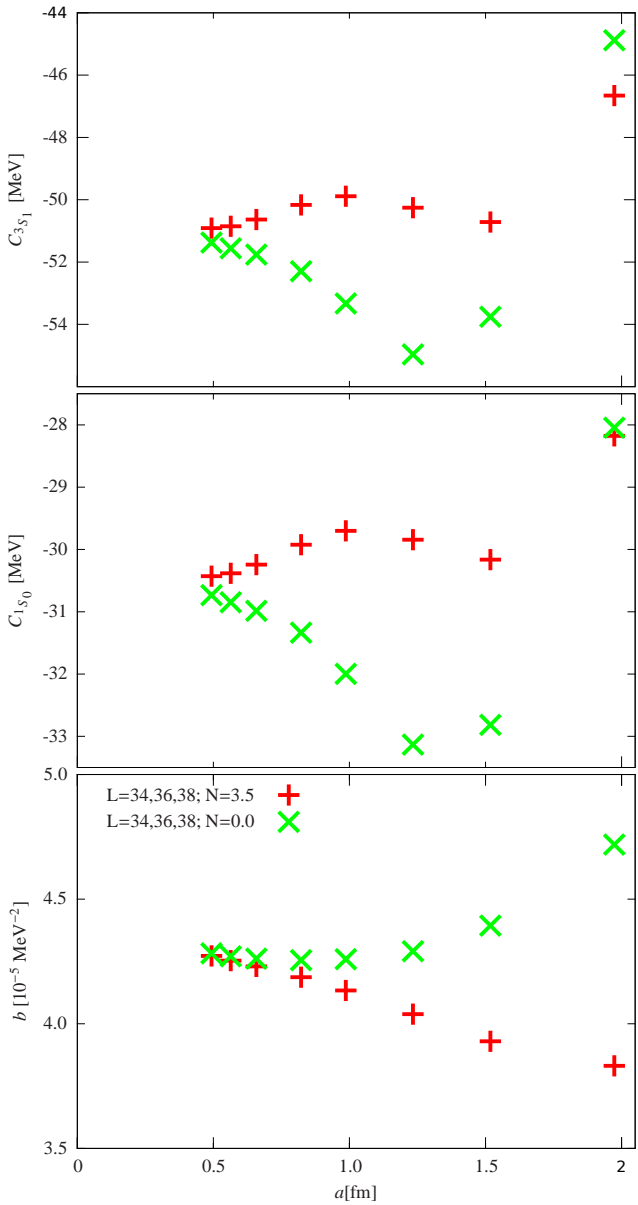}
\caption{Low-energy coupling constants and the smearing parameter $b$ in the pionless EFT for a lattice size of $L=34,36,38$ 
and two different stretching factors.}
\label{fig:LECsmnopion}
\end{figure}

Initially, we consider the pionless case which works well for very low energies and is described by the effective 
Hamiltonian 
\begin{equation}
H=H_{\mathrm{free}}+H_{\mathrm{cont}}~.
\end{equation}
It was shown in Ref.~\cite{Rokash:2013xda} that for any 
lattice spacing the non-smeared contact interaction cannot reproduce the effective range correctly. Hence, we introduce 
a smearing according to Eq.~(\ref{eq:smearing}). The calculation is performed for $N=0$ and $N=3.5$ to estimate the influence of the stretching factor in the improved action. As we only have to 
consider the S-wave projection, we use the appropriate linear combinations
\begin{equation}
C= \frac{1}{4} \left(3C_{^1S_0}+C_{^3S_1}\right) \quad 
C_I= \frac{1}{4} \left(C_{^1S_0}+C_{^3S_1}\right)~~. 
\end{equation}
The results for two stretching factors $N=0$ and $N=3.5$ are shown in Fig.~\ref{fig:LECsmnopion}. For $N=3.5$, the explicit values of the fitted parameters and the predictions are listed in Tab.~\ref{fig:LECsmnopion}. The lattice size of $L=34,36,38$ is motivated by the corresponding physical lattice length of $L_\mathrm{phys}=16.66, 17.64, 18.62$~fm for a minimal lattice spacing of $a=0.49$~fm. This should be still large enough so that higher order terms to Eqs.~(\ref{eq:EBL},\ref{eq:pcotdL}) and in the effective range expansion are negligible. The LECs in the respective table and plot are obtained after rescaling $C_{^3S_1/^1S_0}\rightarrow C_{^3S_1/^1S_0}/a^3$ so as to account for the different volumes.

First, we see the parameters belonging to $N=0$ and $N=3.5$ have similar values and they approach a constant value when the lattice spacing is minimized. But there is a preference for $N=3.5$ case because for this case, the variation of the LECs is significantly decreased for various lattice spacings than in the $N=0$ case reflecting the improved dispersion relation approximation which is necessary as we receive data points in different regimes for different lattice spacings. Therefore, we use $N=3.5$ for the following calculations. 

Comparing the theoretical predictions with the experimental values gives the impression of a large deviation particularly for $r_{^{\,3}S_1}$. As we only work at lowest order, we cannot expect to get the physical value already. But we can perform a consistency check for our predictions as there exists a relation between the binding energy, the scattering length and the effective range, see e.g. Ref.~\cite{Braaten:2004rn}
\be
E_B   
\approx - \frac{1}{2\mu a_s^2}\left(1+\frac{r_e}{a_s}+\frac{5r^2_e}{4a^2_s}+\ldots\right)~.
\label{eq:EBar-relation}
\ee
This equation is valid for positive scattering length $a_s$ and the expansion 
in $r_e/a_s$, with $r_e$ the effective range, is useful provided $\lvert r_e\lvert \ll a_s$. 
As we do not include any 
partial wave mixing, we can use the expansion up to $\mathcal{O}({r_e^2/a_s^2})$ and 
compare the physical binding energy in this scheme with our lattice predictions. 
Then the binding energy is $E_B=-2.052(1)$~MeV for the physical values of the scattering
length and the effective range and our lattice predictions 
give a binding energy of $E_B=-2.009(1)\ldots-1.999(1)$~MeV showing that the 
relation in Eq.~(\ref{eq:EBar-relation}) is approximately fulfilled.

The constant value of $b$ can be explained by the smearing function itself and the Fourier transformation of the contact interaction given in Eq.~(\ref{eq:contIntmod}) in dependence of the cutoff/ lattice spacing. While we work on the lattice, the Fourier transformation is limited by the maximum momentum $\pi/a$. As we consider the continuum limit, new contributions to the summation are added in Eq.~(\ref{eq:contIntmod}) as $\pi/a\rightarrow \infty$ and the value of the normalization constant changes. The normalization constant $f_0$ approaches a finite limit and the new contributions are exponentially suppressed. Hence, the lattice spacing dependence of the smearing constant should approach a constant value as long as other discretization errors are negligible. This is also a good check for other regularizations procedures. We will use this condition to find the best regularization scheme in the upcoming section. It is quite remarkable that the pionless theory can be regularized
by such a constant Gaussian smearing leading to $a$-independent results for $0.5 \lesssim a \lesssim 2.0\,$fm. Given that the typical extension of
a nucleon is given by a scale of $r \simeq 0.85\,$fm, this means that the EFT can be used for all momenta that do not lead to a resolution of the
internal nucleon structure, at least in the pionless theory. A direct comparison with the results of a 
  continuum calculation is difficult because the occuring divergences are usually treated in a different way. 
As done in Ref.~\cite{Phillips:1997xu}, one can calculate 
 the scattering matrix using the bubble chain summation with 
 a regularization $\tilde{f}(\vec{p})=\exp(-b\vec{p}^2/2)$ similar to the smeared contact 
interaction on the lattice instead of dimensional regularization or a finite cut-off.
Then the T-matrix is expanded up to and including $\mathcal{O}(\vec{p}^2)$ and matched to the effective 
range expansion  parameters. The values one obtains are of the same order as the lattice values but 
do not match them exactly. Note further that the extension to the three-particle system is not 
straightforward because of the Efimov
effect that requires the inclusion of a three-body contact interaction already at leading order \cite{Bedaque:1998kg}.

\section{The pionful theory}

Including pions is necessary for an effective field theory at higher energies. 
Therefore pions are included according to Eq.~(\ref{eq:HamOPE}) and the 
full Hamiltonian is 
\begin{equation}
H=H_{\mathrm{free}} +H_{\mathrm{cont}}+H_{\mathrm{OPE}}~.
\end{equation}
A problem is caused by the singularities which arise in the short-range region of the OPE contribution. On the one hand, this singularity exists for any lattice spacing as the relative distance $\vec{r}=0$ is possible and gives a very large but still finite contribution. On the other hand, minimizing the lattice spacing leads to additional lattice points with non-zero but very small distances to the origin and which give an additional large short-range contribution.

Firstly, we switch off the smearing of the contact interaction and include the pion-nucleon interaction according to Eq. (\ref{eq:HamOPE}). In this case, the predictions for the effective range parameter are not close to the physical value and it is still necessary to include the smearing of the contact interaction \cite{Borasoy:2006qn}.

In the case of smeared contact interaction and non-smeared pion-nucleon interaction the LECs do not give a reasonably close value for the effective range in the $^3S_1$-channel. The prediction is reasonable and constant for a lattice spacing larger than 1~fm, but it does decrease towards zero for smaller lattice spacings. As we do not include a regularization, the divergent $\pi$N-contribution for small lattice spacings is more and more resolved for smaller lattice spacings resulting in a very sharp potential. The contact interaction in combination with the smearing factor $b$ can compensate this effect but as the $^1S_0$- and $^3S_1$-channel do not have exactly the same dependence, while it is still possible to fit to one of the effective
ranges, the agreement for the other one gets worse.

Now we turn on the smearing of the pion-nucleon interaction. While a smearing in momentum space according to Eq.~(\ref{eq:smearing}) could be possible, we follow the arguments given in Ref. \cite{Epelbaum:2014efa} and introduce the regularization in position space as proposed in Eq.~(\ref{eq:possmearing}).
Then, the new OPE potential reads
\begin{equation}
\begin{split}
H_{\mathrm{OPE}}=-\frac{g_A^2}{8F_\pi^2}\sum_{S_1,S_2,I}& \sum_{\vec{n}_1,\vec{n}_2}\tilde{f}\left(\vec{n}_1-\vec{n_2}\right)\\ 
& \times G_{S_1S_2}\left(\vec{n}_1-\vec{n}_2\right)\rho_{S_1,I}^{a^\dagger,a}\left(\vec{n}_1\right) \rho_{S_2,I}^{a^\dagger,a}\left(\vec{n}_2\right)~.
\end{split}
\label{eq:HamOPEmod}
\end{equation}
\begin{figure}[htb]
\psfrag{L}[][]{\rotatebox{180}{\scriptsize$L$~[lattice units]}}
\psfrag{ pcot}[][]{\rotatebox{180}{\scriptsize $p\cot\left(\delta\right)$~[MeV]}}
\psfrag{ p^2[MeV^-2]}[][]{\scriptsize$p^2$~[MeV$^{2}$]}
\psfrag{L=141720pos}[][]{\scriptsize L=14,17,20}
\psfrag{L=343638pos}[][]{\scriptsize L=34,36,38}
\psfrag{L=384042pos}[][]{\scriptsize L=38,40,42}
\psfrag{ E_b[MeV]}[][]{\rotatebox{180}{\scriptsize$E_B$~[MeV]}}
\psfrag{0.0}{\scriptsize 0}
\psfrag{-2.20}{\scriptsize -2.20}
\psfrag{-2.25}{\scriptsize -2.25}
\psfrag{-2.30}{\scriptsize -2.30}
\psfrag{-2.35}{\scriptsize -2.35}
\psfrag{-2.40}{\scriptsize -2.40}
\psfrag{-2.45}{\scriptsize -2.45}
\psfrag{-2.50}{\scriptsize -2.50}
\psfrag{34}{\scriptsize 34}
\psfrag{36}{\scriptsize 36}
\psfrag{38}{\scriptsize 38}
\psfrag{9}{\scriptsize 9}
\psfrag{8}{\scriptsize 8}
\psfrag{7}{\scriptsize 7}
\psfrag{6}{\scriptsize 6}
\psfrag{5}{\scriptsize 5}
\psfrag{-500}{\scriptsize -500}
\psfrag{-400}{\scriptsize -400}
\psfrag{-300}{\scriptsize -300}
\psfrag{-200}{\scriptsize -200}
\psfrag{-100}{\scriptsize -100}
\psfrag{0}{\scriptsize 0}
\psfrag{500}{\scriptsize 500}
\psfrag{1000}{\scriptsize 1000}
\psfrag{1500}{\scriptsize 1500}
\psfrag{a=1.97fm}[][]{\scriptsize a=1.97~fm}
\psfrag{a=0.82fm}[][]{\scriptsize a=0.82~fm}
\psfrag{a=0.66fm}[][]{\scriptsize a=0.66~fm}
\psfrag{a=0.56fm}[][]{\scriptsize a=0.56~fm}
\psfrag{a=0.49fm}[][]{\scriptsize a=0.50~fm}
\includegraphics[width=0.48\textwidth]{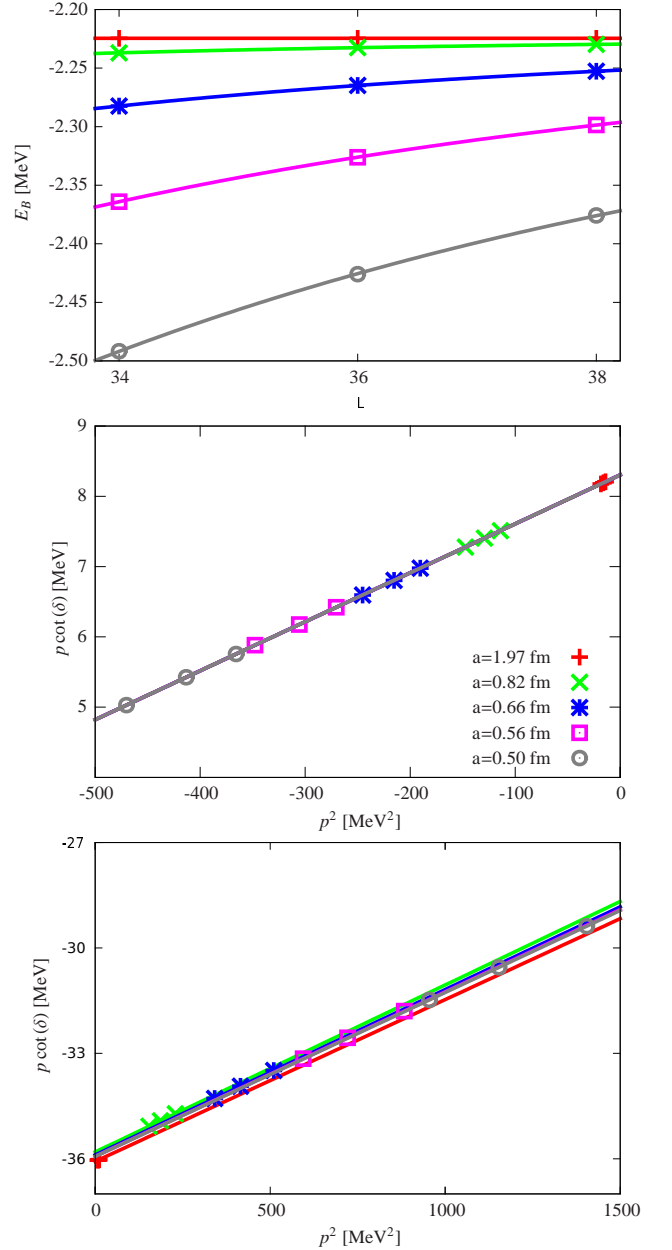}
\psfrag{-5.0}{\scriptsize -0.5}
\caption{Lattice space dependence of various parameters. Upper panel: The finite-volume deuteron bound state energy. Central panel: The effective range expansion for the $^1S_0$-channel. Lower panel: The effective range expansion for the $^3S_1$-channel.}
\label{fig:EREpionEE}
\end{figure}

\noindent Throughout this paper, we use $n=4$ in Eq.~(\ref{eq:possmearing}), but we also checked for different $n$ and the results are similar. Higher values of $n$ are only necessary for strongly divergent contributions like two-pion-exchange potentials which we neglect in this exploratory study.\\
\noindent  In Fig.~\ref{fig:EREpionEE}, the finite volume binding energy of the deuteron and the respective effective range expansion for the S-waves for a subset of the lattice spacings used are shown. Finite volume effects modify the binding energy so that a correction according to Eq. (\ref{eq:EBL}) is necessary but higher order terms are still negligible. Also the data points we obtain for the effective range expansion still have small enough momenta squared so that the expansion up to $\mathcal{O}(p^2)$ is feasible. 
The resulting LECs, the smearing constant as well as the predictions for $L= 34,36,38$ are displayed in Tab.~\ref{tab:LECspionsscposmpi}. While the value of the LECs are different to the pionless case, their general shape depending on the lattice spacing does not change and the LECs remain negative as well. Also the smearing factor $b$ remains in a certain range between 
$2.40$ to $2.92$~$\times 10^{-5}\,$MeV$^{-2}$, indicating that this regularization scheme works quite well. We do not show the results for the smaller set of lattice sizes, but we have shown that finite volume effects between these two sets become negligible for lattice spacings larger than 0.7~fm. Assuming a large enough lattice there is a rise in the smearing constant between 1.8~fm and 0.7~fm, and it is constant again for lattice spaces smaller than 0.7~fm. While in the range from 0.7~fm and 1.8~fm the pion-nucleon contribution changes, the lattice sites closest to the origin contribute more and more to the potential due to the divergent structure of the potential. This effect is compensated by the regularization at a certain range as we further decrease the lattice spacing, and then the regularized potential does not change its shape anymore.
Note that in the intermediate range there are lattice artifacts in the OPEP which cause the oscillatory behavior of the LECs as they appear at different physical lengths and multiplied by a different regularization factor. As a result, we cannot see a plateau for the LECs $C_{^3S_1}$ and $C_{^1S_0}$ at a lattice spacing of $a=0.5$~fm, but we have to further decrease the lattice spacing. Simultaneously, we increase the number of lattice points to keep the physical lattice size. The respective values are shown in 
Tab.~\ref{tab:LECspionsscposmpi} as well and there is a plateau starting from $a=0.39$~fm. The predictions for the effective range expansion parameters are also quite good as $a_{^{\,3}S_1}\approx 5.470(1) \ldots 5.649(1)$~fm and $r_{^{\,3}S_1}\approx 1.818(1) \ldots 1.899(1)$~fm depending on the lattice spacing. Repeating the calculation for the binding energy according to  Eq.~(\ref{eq:EBar-relation}) up-to-and-including terms of $\mathcal{O}({r_e^2/a_s^2})$
yields for the binding energy $E_B\approx 2.026(1) \ldots 2.033(1)$~MeV for the various 
lattice spacings in the range between 0.5~fm and 2.0~fm. This is again close to the physical binding energy and the respective 
relation is fulfilled. For the remaining ones, it yields $E_B\approx 2.034(1) \ldots 1.899(1)$~MeV indicating that larger volumes become necessary particularly for the smallest lattice spacing. Finally, the theory is well regularized and the differences 
between the lattice predictions and the experimental values of $a_{^{\,3}S_1}$ and $r_{^{\,3}S_1}$, 
respectively, can be compensated by higher order terms.
\renewcommand{\arraystretch}{1.05}
\begin{SCtable*}[25][!hbt]
\centering
\begin{tabular}{|c|c|c|c|c|c|c|}
\hline
$a^{-1}$~[MeV]&$a$~[fm]&$C_{^3S_1}$~[MeV]&$C_{^1S_0}$~[MeV]&$b$~[MeV$^{-2}$]& $a_{^{\,3}S_1}$~[fm]&$r_{^{\,3}S_1}$~[fm]\\
\hline
100	&1.97& $-54.07$&$-36.11$&$2.59\cdot10^{-5}$&$5.470(1)$&$1.818(1)$\\
130	&1.52& $-67.11$&$-45.47$&$2.40\cdot10^{-5}$&$5.513(1)$&$1.878(1)$\\
160	&1.23& $-69.31$&$-46.52$&$2.41\cdot10^{-5}$&$5.527(1)$&$1.899(1)$\\
200	&0.99& $-63.83$&$-41.39$&$2.65\cdot10^{-5}$&$5.523(1)$&$1.893(1)$\\
240	&0.82& $-61.55$&$-39.23$&$2.83\cdot10^{-5}$&$5.511(1)$&$1.876(1)$\\
300	&0.66& $-62.20$&$-39.44$&$2.92\cdot10^{-5}$&$5.498(1)$&$1.858(1)$\\
350	&0.56& $-63.30 $&$-40.19$&$2.92\cdot10^{-5}$&$5.491(1)$&$1.842(1)$\\
400	&0.49& $-64.31$&$-40.97$&$2.91\cdot10^{-5}$&$5.491(1)$&$1.842(1)$\\
\hline
500 &0.39& $-65.47$&$-41.90$&$2.89\cdot10^{-5}$&$5.500(1)$&$1.836(1)$\\
600 &0.33& $-66.14$&$-42.71$&$2.82\cdot10^{-5}$&$5.553(1)$&$1.835(1)$\\
700 &0.28& $-65.61$&$-42.64$ &$2.86\cdot10^{-5}$&$5.649(1)$&$1.845(1)$\\
\hline
\end{tabular}
\caption{Fit results in EFT with Gaussian-smeared contact interaction and position space regularization for the pion-nucleon interaction with $N=3.5$. The Lattice size is $L=34,36,38$ for $a= 0.49 \ldots 1.97$ ~fm and $L=38,40,42$ for $a= 0.39 \ldots 0.28$~fm. The values for $a_{^{\,3}S_1}$ and $r_{^{\,3}S_1}$ are predictions
(modulo the consistency condition Eq.~(\ref{eq:EBar-relation})).}
\label{tab:LECspionsscposmpi}
\end{SCtable*}
\renewcommand{\arraystretch}{1.0}


\section{Conclusion and Outlook}

In this paper, we have studied the lattice spacing $a$ dependence of the neutron-proton system at leading order in the pionless and the pionful theory.
We have used the scattering length and the effective range in the $^1S_0$ partial wave together with the deuteron binding energy to fix the two LECs
related to the LO four-nucleon contact interaction and the smearing para\-meter $b$. To focus on the lattice spacing dependence, we have worked at
sufficiently large lattice volumes so that finite volume effects do not play any role here.
In the pionless case, it is sufficient to smear the contact interactions with a Gaussian-type function, cf. Eq.~(\ref{eq:smearing}), to achieve $a$-independence
in the range $0.5 \lesssim a \lesssim 2.0\,$fm. We have explicitly shown this for the scattering length and the effective range in the $^3S_1$-channel, being
aware of the strong correlation between $E_B$ and $a_{^{\,3}S_1}$.
In the pionful theory, the contribution from the one-pion-exchange is best to be regularized in position space, as discussed in detail in Ref.~\cite{Epelbaum:2014efa}.
Again, we can demonstrate lattice spacing independence for the same range of $a$. Therefore, it should be possible to calculate the phase shifts with the transfer matrix method and the spherical wall method. Smaller lattice spacings should minimize the lattice errors such as the broken rotational invariance and make it possible to increase the accuracy especially for higher momenta in the partial wave decomposition. Clearly, when decreasing the lattice spacing, one has to be aware that the perturbative
treatment of the NLO and higher order corrections becomes doubtful and requires a separate investigation but it does not 
invalidate the results found here. Furhermore, one should implement this regularization in Monte Carlo simulations in order to minimize the lattice spacing and artefacts in the simulation. This, however, is a separate issue.

\section*{Acknowledgments}
We thank J.~M.~Alarcon, S.~Bour, S.~Elhatisari, \mbox{E.~Epelbaum} and H.-W.~Hammer for discussion.
We acknowledge partial financial support from the Deutsche Forschungsgemeinschaft (Sino-German CRC 110).

\bibliographystyle{model1-num-names.bst}

\end{document}